\documentclass[12pt]{article}
\usepackage{geometry}                
\usepackage{graphicx}
\usepackage{amssymb}
\usepackage{amsmath}
\usepackage{epstopdf}

\usepackage[hyperindex=true,
          pdfstartview=FitH,
          bookmarksnumbered=true,
          bookmarksopen=true,
          citecolor=blue,
          linkcolor=blue,
          colorlinks=true,
          pdfborder=001,
          unicode]{hyperref}

\begin{document}

\title{Note on a class of anisotropic Einstein metrics}
\author{Liu Zhao\\
School of Physics, Nankai university, Tianjin 300071, China\\
email: lzhao@nankai.edu.cn}
\date{}                                           
\maketitle

\begin{abstract}
A class of anisotropic Einstein metrics is presented. These metrics are axial symmetric 
and contains an anisotropy parameter $\alpha$, which is 
identified as the amplitude of the proper acceleration of the origin, thus explaining the 
ellipsoid shape of the horizon. 
\end{abstract}

Among exact solutions to Einstein equation, the constant curvature solutions, i.e. 
Einstein metrics, play a very special role, because such solutions correspond to 
gravitational fields with no matter distribution, which in physics terms dabbed ``vacuum 
solutions''.  In modern cosmological context, the positive constant curvature spacetime
(i.e. de Sitter spacetime) is of particular importance, as it is regarded to be the future 
asymptotics of our universe. Mathematically, the constant curvature 
(pseudo-)Riemannian metrics are also of great importance, because they represent the  
maximally symmetric spaces and are often used as of ideal models of more complicated 
curved spaces.

The prototype Einstein metrics with positive, zero and negative constant curvatures are 
given respectively by the de Sitter, Minkowski and anti-de Sitter line elements
\begin{align}
ds_{\mathrm{dS}}^{2} &= -\left(1-\frac{r^{2}}{\ell^{2}}\right)dt^{2}
+\left(1-\frac{r^{2}}{\ell^{2}}\right)^{-1}dr^{2} + r^{2} d\Omega_{n-2}^{2},
\label{dS}\\
ds_{\mathrm{M}}^{2} &= -dt^{2}
+dr^{2} + r^{2} d\Omega_{n-2}^{2},\\
ds_{\mathrm{AdS}}^{2} &= -\left(1+\frac{r^{2}}{\ell^{2}}\right)dt^{2}
+\left(1+\frac{r^{2}}{\ell^{2}}\right)^{-1}dr^{2} + r^{2} d\Omega_{n-2}^{2},
\end{align}
where $n$ is the spacetime dimension and $d\Omega_{n-2}^{2}$ is the line element for a 
unit $(n-2)$-sphere
\[
d\Omega_{n-2}^{2}=d\theta_{1}^{2}+\sin^{2}\theta_{1}(d\theta_{2}^{2}+
\sin^{2}\theta_{2}(d\theta_{3}^{2}+...)).
\]
We denote $\theta_{n-2}$ as $\phi$. The parameter $\ell$ is called the (A)dS radius and is related to the 
cosmological constant $\Lambda$ via
\[
\ell^{2} = \pm \frac{(n-1)(n-2)}{2\Lambda}.
\] 
These line elements are explicitly isotropic about the origin 
and it can be shown without much effort that they are also homogeneous. Actually, as 
pointed out in \cite{Wald:1984un}, a maximally 
symmetric space is necessarily homogeneous and isotropic about all points. 

Naturally one can raise questions such as ``are there any Einstein metric which is 
anisotropic or non-homogeneous?'' The answer is certainly yes. Actually, non-
homogeneous Einstein metrics are not rare. Any vacuum black hole solution (e.g. 
Schwarzschild and Schwarzschild-(A)dS black holes) will break the homogeneity whilst 
still preserves the isotropy about the origin.  Vacuum black hole solutions with rotation 
parameters (e.g. Kerr and Meyers-Perry black holes) will also break the isotropy about 
the origin. So the answer to the above question seems to be well known. However, 
in the absence of black holes, examples of Einstein metrics which are anisotropic or 
non-homogeneous are less well known, with the exceptions of warped embedded 
spaces, see, e.g. \cite{Kim:2003p123}\cite{Yang:2010ut}\cite{Petersen:2010p121}\cite{Petersen:2010p120}. The present note is aimed to present a novel 
class of anisotropic Einstein metrics beyond the class of warped embeddings which can 
still be presented in a very neat and compact manner. 

The novel metrics which we have in mind are given as follows:
\begin{align}
ds_{1}^{2} &= \omega(r,\theta_{1})^{2}\left[-\left(1-\frac{r^{2}}{\ell^{2}}\right)dt^{2}
+\left(1-\frac{r^{2}}{\ell^{2}}\right)^{-1}dr^{2} + r^{2} d\Omega_{n-2}^{2}\right],\label{1} \\
ds_{2}^{2} &= \omega(r,\theta_{1})^{2}\left[-dt^{2}
+dr^{2} + r^{2} d\Omega_{n-2}^{2}\right],\label{2}\\
ds_{3}^{2} &= \omega(r,\theta_{1})^{2}\left[-\left(1+\frac{r^{2}}{\ell^{2}}\right)dt^{2}
+\left(1+\frac{r^{2}}{\ell^{2}}\right)^{-1}dr^{2} + r^{2} d\Omega_{n-2}^{2}\right],\label{3}
\end{align}
where the overall conformal factor $\omega(r,\theta)$ is given by
\begin{align}
\omega(r,\theta)=\frac{1}{1-\alpha r \cos(\theta)}. \label{omega}
\end{align}
The explicit dependence of $\omega(r,\theta)$ on $\theta$ breaks the isotropy of the 
original dS, Minkowski and AdS metrics. Each of the 3 metrics depends on the 
parameter $\alpha$ through $\omega(r,\theta)$, while (\ref{1}) and (\ref{3}) depend 
on an extra parameter $\ell$. Since at $\alpha=0$, $\omega(r,\theta)=1$, we see that 
the anisotropy of the above metrics are solely caused by the presence of the 
parameter $\alpha$, and we may hence refer to it as the anisotropy parameter. 
Each of the 3 metrics satisfies the vacuum 
Einstein equation $R^{(i)}_{\mu\nu}-\frac{1}{2}R^{(i)}g^{(i)}_{\mu\nu}+
\Lambda_{i} g^{(i)}_{\mu\nu}=0$ with $i=1,2,3$, and the free parameters are 
explicitly related to cosmological constant via
\begin{align}
\Lambda_{1}&=\frac{1}{2}(n-1)(n-2)\left(\frac{1}{\ell^{2}}-\alpha^{2}\right),
\label{lambda}\\
\Lambda_{2}&=-\frac{1}{2}(n-1)(n-2)\alpha^{2},\\
\Lambda_{3}&=-\frac{1}{2}(n-1)(n-2)\left(\frac{1}{\ell^{2}}+\alpha^{2}\right).
\end{align}
We can easily see that the metric (\ref{2}) can be seen as the $\ell\rightarrow \infty$ 
limit of either (\ref{1}) or (\ref{3}), whilst (\ref{3}) can be obtained via $\ell 
\rightarrow i\ell$ from (\ref{1}). Therefore, we need only to consider the metric (\ref
{1}) in the following. We henceforth drop the suffix 1 from the line element (\ref{1}) 
and the corresponding cosmological constant (\ref{lambda}). We shall consider only 
the cases with $n>2$.

The cosmological constant $\Lambda$ in (\ref{lambda}) can take positive, zero and 
negative values depending on the values of $\alpha$ and $\ell$ (we assume both $
\alpha$ and $\ell$ are positive):
\begin{align*}
\Lambda >0 &\quad\mbox{if}\quad \alpha\ell<1,\\
\Lambda =0 &\quad\mbox{if}\quad \alpha\ell=1,\\
\Lambda <0 &\quad\mbox{if}\quad \alpha\ell>1.
\end{align*}
We thus see that the geometry of the metric described by the line element (\ref{1}) 
differs drastically from the de Sitter metric (\ref{dS}), even though they are simply 
conformally related to each other. A straightforward consequence of this difference 
concerns the horizon structure. For the de Sitter metric (\ref{dS}), $r=\ell$ represents 
a cosmological horizon. However, for the metric (\ref{1}), things differ. If $\alpha
\ell>1$, conformal infinities will be reached before $r$ approaches the value $\ell$, 
thus $r=\ell$ is beyond the coordinate patch of the spacetime described by the chosen 
coordinate system. If  $\alpha\ell=1$, $r$ can approach the value $\ell$, but this 
constant $r$ hypersurface contains the conformal infinity in some direction and 
so does not describe a complete horizon. If $\alpha\ell<1$, $r$ can approach the 
value $\ell$ without touching conformal infinities in any direction, and hence $r=\ell$ 
represents a complete horizon. To fully describe the horizon structure we will have to 
find global coordinates corresponding to the spacetime (\ref{1}), which is a 
complicated task and is beyond the scope of this note. We emphasis that hitting the 
conformal infinity before reaching the $r=\ell$ hypersurface does not imply that 
horizon do not exist. It just means that the coordinate system which we use will 
breakdown at the region which is under exploration.  For a detailed discussion on the 
global structure of a similar spacetime in 4D and 5D, see \cite{Dias:2002}, 
\cite{Dias:2003} and \cite{Zhao-Xu}\footnote{The massless uncharged 
C-metric spacetimes in \cite{Dias:2002}{Dias:2003} is very similar to the 4D 
version of (\ref{1}) and (\ref{3}), with only a slightly different definition of the overall 
conformal factor. The same similarity happens between the spacetime discussed in 
\cite{Zhao-Xu} and the 5D version of (\ref{1}).}.

\begin{figure}[ht]
\begin{center}
\includegraphics[height=8cm]{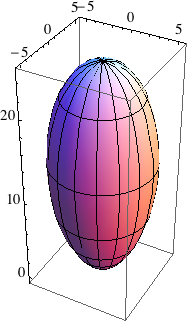}
\end{center}
\caption{Plot of the horizon surface for $n=4$ with $\alpha\ell=0.8$}
\end{figure}

In spite of the sharp differences of the spacetime (\ref{1}) from de Sitter spacetime 
(\ref{dS}), we can still find some common features of the two metrics. The most 
significant common feature is the following: every light-like vector for the spacetime 
(\ref{dS}) will remain to be light-like in (\ref{1}). Therefore, in the case when 
$r=\ell$ can be approached and describes a horizon, we would ask what the geometry 
of the horizon looks like in the spacetime (\ref{1}). This horizon geometry is described 
by the line element
\[
ds_{H}^{2}=\omega(\ell,\theta_{1})^{2}\ell^{2}d\Omega_{n-2}^{2}.
\]
Clearly this is a conformal $(n-2)$-sphere but we would like to have a vivid view on 
the actual shape. For the most interesting case of $n=4$ we can make a plot of the 
surface, as shown in Fig.1.
This is an ellipsoid with the $\theta_{1}=0$ direction playing the role of a rotational 
symmetric axis. This is not a surprise because the function $\omega(r,\theta)$ for 
fixed $r$ obeying $\alpha r <1$ represents an ellipse and this plays the role of a 
variable radius for the horizon surface at any constant $\phi$.

It is tempting to ask what makes the horizon looks like an ellipsoid. Bearing in mind 
that the horizon(s) may exist even when the cosmological constant is zero or negative, 
it is not difficult to guess that such horizons can only be caused by acceleration. It is 
indeed so. To clarify this point, we take a static observer following a timelike geodesics 
in the spacetime (\ref{1}). Such an observer is represented by the worldline $x^\mu
(\tau) = (\omega(r,\theta_{1})^{-1} \ell \tau/
\sqrt{\ell^2-r^2}, r, \theta_{1},\theta_{2},...,\phi)$, where $\tau$ is the proper time 
and $r, \theta_{1},\theta_{2},...,\phi$ are independent of $\tau$. The proper velocity 
is given by
\[
u^{\mu}=(\omega(r,\theta_{1})^{-1} \ell /\sqrt{\ell^2-r^2},0,...,0).
\]
Hence, the proper acceleration $a^{\mu}=u^{\nu}\nabla_{\nu}u^{\mu}$ obeys
\begin{align}
 a^\mu a_\mu&=\alpha^{2} - \frac{1}{\ell^{2}} 
  + \frac { (1 - \alpha r\cos\theta_1)^{2}}{ \ell^{2} - r^{2} }\nonumber\\
  &=\frac { (1 - \alpha r\cos\theta_1)^{2}}{ \ell^{2} - r^{2} }
  -\frac{2\Lambda}{(n-1)(n-2)}. \label{amp}
\end{align}
At $r=0$, the amplitude of the proper acceleration is $|a|=\alpha$, which gives a 
geometric explanation for the meaning of the parameter $\alpha$.

It is of particular interest to understand the case $n=4$ more thoroughly, because this 
case may have some cosmological implications. The proper acceleration 
amplitude (\ref{amp}) indicates that $\theta_{1}=0$ and $\theta_{1}=\pi$ are two 
extrema of the proper acceleration. This fact, combined with the ellipsoid shape of the 
horizon shown in Fig.1, indicates that there should be a dipole moment from the point 
of view of the accelerating observer. This dipole effect reminds us of the CMB dipole 
moment. The standard explanation, the CMB dipole moment is due to the peculiar 
motion of our milky way with respect to the nearby galaxy cluster. We should now ask 
whether this peculiar motion contains some acceleration factor. Moreover, we can 
infer from (\ref{amp}) that even when $\Lambda=0$, the proper acceleration does not 
vanish, and thus according to Unruh effect, there will be an acceleration horizon 
appearing in the spacetime from the point of view of the accelerating observer. Is the 
observed acceleration of the universe solely due to the proper acceleration of ourselves? 

\section*{Acknowledgment} 

This work is supported by the National  Natural Science Foundation of 
China (NSFC) through grant No.10875059. The author would like to thank the 
organizer and participants of ``The advanced workshop on Dark Energy 
and Fundamental Theory'' supported by the Special Fund for Theoretical 
Physics from the National Natural Science Foundation of China with grant 
no: 10947203 for comments and discussions.

\providecommand{\href}[2]{#2}\begingroup
\begin
{thebibliography}{13}

\bibitem{Wald:1984un} S. Weinberg, ``Gravitation and Cosmology: Principles and Applications of the General Theory of Relativity'', (John Wiley \& Sons Inc, 1972), p. 379.

\bibitem{Kim:2003p123}D. Kim, Y. H. Kim, ``Compact Einstein warped product spaces with nonpositive scalar curvature'', Proceedings of the American Mathematical Society (2003) vol. 131 (8) pp. 2573-2576.

\bibitem{Yang:2010ut} H.-X. Yang, L. Zhao,  ``Warped embeddings between Einstein manifolds'', Mod. Phys. Lett. A 25, No. 18 (2010) pp. 1521-1530 [{arXiv}
{\bf hep-th} (Feb, 2010)  \href{http://www.arXiv.org/abs/1002.1001v1}
{{\tt 1002.1001v1}}].

\bibitem{Petersen:2010p121}C. He, P. Petersen and W. Wylie, ``On the classification of warped product Einstein metrics'', {arXiv}
{\bf math.DG} (Oct, 2010)  \href{http://www.arXiv.org/abs/1010.5488v1}
{{\tt 1010.5488v1}}.

\bibitem{Petersen:2010p120}C. He, P. Petersen and W. Wylie, ``Warped product Einstein metrics over spaces with constant scalar  curvature'', {arXiv}
{\bf math.DG} (Dec, 2010)  \href{http://www.arXiv.org/abs/1012.3446v1}
{{\tt 1012.3446v1}}.

\bibitem{Dias:2002}
O.~J.~C. Dias and J.~P.~S. Lemos, ``Pair of accelerated black holes
in an anti-de Sitter background: the AdS C-metric,'' Phys.Rev. D67
(2003)  064001 [{arXiv:} {\bf hep-th} (Mar, 2003)
\href{http://www.arXiv.org/abs/hep-th/0210065v3}{{\tt hep-th/
0210065v3}}].

\bibitem{Dias:2003}
O.~J.~C. Dias and J.~P.~S. Lemos, ``Pair of accelerated black holes
in a de  Sitter background: the dS C-metric,'' Phys. Rev. D67 (2003)
084018 [{arXiv:} {\bf hep-th} (Jan, 2003)
\href{http://www.arXiv.org/abs/hep-th/0301046v2}{{\tt hep-th/
0301046v2}}].

\bibitem{Zhao-Xu}W. Xu, L. Zhao, and B. Zhu, ``Five-dimensional vacuum Einstein 
spacetimes in C-metric like coordinates'', Mod. Phys. Lett. A, Vol. 25, No. 32 (2010) pp. 2727-2743 [{arXiv}
{\bf hep-th} (Feb, 2010)  \href{http://www.arXiv.org/abs/1005.2444v1}
{{\tt 1005.2444v3}}].

\end{thebibliography}\endgroup

\end{document}